# Construction and Application of Teaching System Based on Crowdsourcing Knowledge Graph

Jinta Weng[1][0000-0002-6235-6454] and Ying Gao[2*] and Jing Qiu[2] and Guozhu Ding[3] and Huanqin Zheng[2]

[1] Guangzhou University, School of Computer Science and Cyber Engineering. CHINA
552122632@qq.com
[2*] Guangzhou University, School of Computer Science and Cyber Engineering. CHINA
falcongao@sina.com.cn
[3] Guangzhou University, School of Education. CHINA

**Abstract.** [Objective] Through the combination of crowdsourcing knowledge graph and teaching system, research methods to generate knowledge graph and its applications. [Method]Using two crowdsourcing approaches, crowdsourcing task distribution and reverse captcha generation, to construct knowledge graph in the field of teaching system. [Results] Generating a complete hierarchical knowledge graph of the teaching domain by nodes of school, student, teacher, course, knowledge point and exercise type. [Limitations] The knowledge graph constructed in a crowdsourcing manner requires many users to participate collaboratively with fully consideration of teachers' guidance and users' mobilization issues. [Conclusion] Based on the three subgraphs of knowledge graph, prominent teacher, student learning situation and suitable learning route could be visualized. [Application] Personalized exercises recommendation model is used to formulate the personalized exercise by algorithm based on the knowledge graph. Collaborative creation model is developed to realize the crowdsourcing construction mechanism. [Evaluation] Though unfamiliarity with the learning mode of knowledge graph and learners' less attention to the knowledge structure, system based on Crowdsourcing Knowledge Graph can still get high acceptance around students and teachers.

**Keywords:** Educational Knowledge Graph; learning analysis; information extraction; Semantic Network; Crowdsourcing

## 1    Introduction

Proposed by Google in 2012,Knowledge graph is a new dynamically correlated knowledge representation, which represent entity and entity' relationship through nodes and links. With the help of knowledge graph, the association and visualization of knowledge will be friendly presented.

Generally, human's cognition of teaching knowledge is always based on our culture and history, and this knowledge is always closely related to its corresponding author, object, history, emotion and other entities. We use the Chinese educational knowledge in the Mu Du's poem, "the southern dynasty four hundred and eighty temple, how many buildings in the misty rain" as an example, and then chooses manual way to get the following entities:

(1) The author is Mu Du.

(2) there are temples and rain.

(3) The dynasty was tang dynasty.

(4) The emotion of ancient poetry is the beauty of Jiang Nan.

By (1)(2)(3)(4) in series four short sentences, it is not difficult to draw, because the author Mu Du live in the ancient millennium, the poem use "temples" "rain" such nice hazy type of entity to represent "the beauty of Jiang Nan".

With Chinese "Education informatization 2.0 Action Plan" pointing out [2]: it is necessary to change the concept of education resource and turn it into continuous and connective view of 'Big Source'. The view of Big Resource means that we should reveal the mobility and interconnection between knowledge and data, and treat educational objects in an interdisciplinary, dynamic and connectable way.

Knowledge Graph subsequently become the popular way to represent educational knowledge. Wentao Hu tracked the relevant data in the learning process of learners and drew the learning path diagram through knowledge graph [3]. Zhen Zhu classified English exercises and generated knowledge graph of English exercises for intelligent topic selection [4]. Zhun Kang and Dejun Wang generated a large biological knowledge base knowledge graph, and constructed an intelligent system for answering questions [5]. However, relationships among different disciplines, courses, knowledge and resources is often neglected in current KG educational systems, resulting in low interoperability and knowledge reusability of educational data. Secondly, crowdsourcing-construction method can also realize in a large degree by massive students and teachers' efforts. Therefore, this paper will introduce and design two crowdsourcing methods to develop a way to general educational knowledge graph.

## 2  Motivation Of Crowdsourcing

### 2.1 General Construction Of KG

Construction of Knowledge graph is information extraction matter. It can be dividing into NER (Named entity recognition) step and RE (Relation extraction) step. Entity annotation, also known as Named Entity Recognition. Relationship extraction，through a certain method, such as TransE[6],, to generate the relationship between the different entities. Both extracting methods are consist of automatic construction method and manual annotation method.

1) Automated construction.

Automated construction method can be divided into two types: Feature Matching and Classification. Feature Matching means full consideration to self-feature, contextual feature and self-restriction of the type of the specified word. Classification method mainly includes HMM proposed by D sahara et al. [7], decision tree model proposed by S sahara [8], SVM model proposed by m. Sahara and Matsumoto [9] and distance-based model TransE. Classification also contains propose appropriate neural network model for deep learning, such as CNN, RNN, and CRF+LSTM method. Thought out the probably existing problem of polysemy in the process of automated construction, word2vec, TF-IDF model, BM25 model, proximity model and semantic feature model facilitate Entity disambiguation and relation disambiguation.

It can be seen that in the field of automatic annotation, whether it is entity annotation or knowledge representation learning, the selected model not only relies heavily on the existing data type and data amount, but also has low accuracy for atlas data with rapid knowledge update and change.

2) Manually Annotated

Manual annotation method means that developers can arrange a certain number of non-user groups to develop and implement the data graph according to certain input specifications. For example, Ruoping Tian et al. used manual annotation to optimize video retrieval [12].This method is often used to build query manuals and professional dictionaries. Because data entry personnel are generally intelligent, the knowledge graph generated can reflect specialization and standardization, but it cannot reflect the diversity and complexity of individual knowledge of human brain.

Although these two methods can solve the problem of knowledge system construction to a certain extent, there are also problems of large expensive manual assistance and high demand for data.

**2.2 Crowdsourcing Within Students' Effort**

Crowdsourcing results have been certificated better than each of individual annotations in swarm intelligence task like group collaboration [13]. Surowiecki found crowdsourcing translation is far better than a single machine translation in low error rate [14].

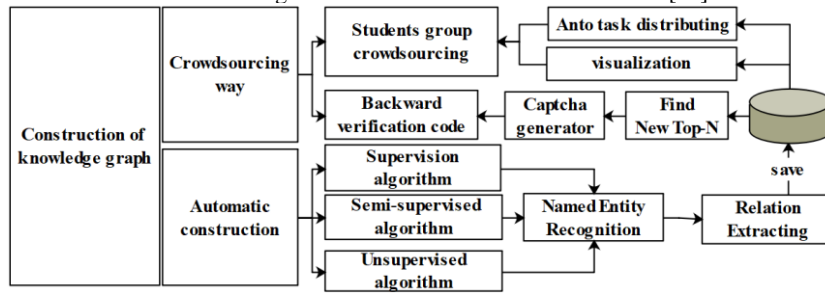

**Fig. 1.** Our construction approach of KG

With the combination of crowdsourcing way and automatic construction, this paper purpose students' group crowdsourcing [15] and backward verification code to use crowdsourcing. As shown in figure 1, different supervised algorithm to realize Named entity recognition, and then use TransE, CNN+LSTM method or other RE algorithm within desirable entities to generate relations which will subsequently save in the triple database. In the part of crowdsourcing, we adopt two methods to generate more triples. First, we design an auto task distributing method to dispense constant type of triples in the corresponding group and develop a visualization tool that enable user directly edit or delete triples. Especially, we use the backward verification code by captcha generator from the Top-N triple.

### 2.3 Crowdsource Task Distribution

As figure 2 shown, at the beginning of the crowdsourcing task module, users will be authenticated and divided into three types: common users, group users and system administrators. For common users, the system will judge whether the user has joined the group. User will accept the task distributed by the group administrator and complete corresponding task after join the group.

The types of tasks can be one of triple verification, perfection of concept and attribute, relationship expansion. Every time common users complete a task, they will automatically trigger the system's reward mechanism and obtain a certain score. Since the wisdom of the group is superposed by the wisdom of the individual, the group score will also be increased and then affect the latter group task proportion.

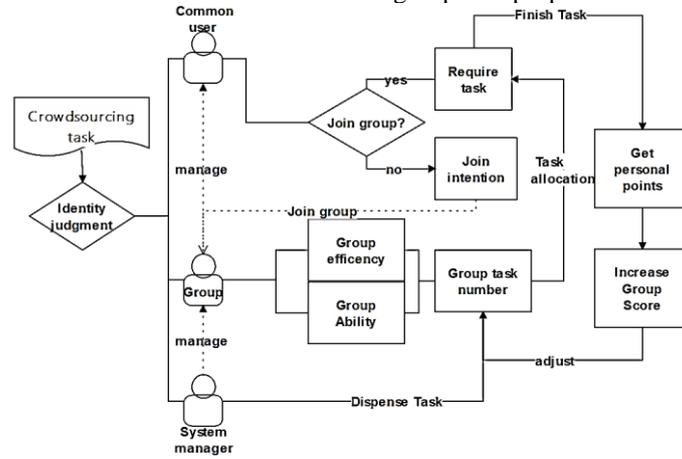

**Fig. 2.** Our construction approach of KG

For the group manager, the group manager is responsible for the task arrangement of the group members. If the group administrator sets the group purpose for the improvement of "Chinese tang dynasty poets and their verses", he/she will receive the system's automatic allocation of the relevant knowledge graph triples. For senior system administrators, they have the authority to manage ordinary users and group administrators. Group administrators can directly manage all group administrators and ordinary users, including group dissolution, group task assignment and group addition.

### 2.4 Reverse Captcha Generation

Reverse captcha generation is mainly reflected in the generation of captchas in the system login page, so captchas generated by captchas generator will extract triples existing in the database and then pack into Full-in-blank type or confirmatory question.

As figure 3 depicted, owing to knowledge grape is make up of entities and links. Question can be divided into entity concept catechism; entity attributes question or relations judgment problem like type 1 to 3.

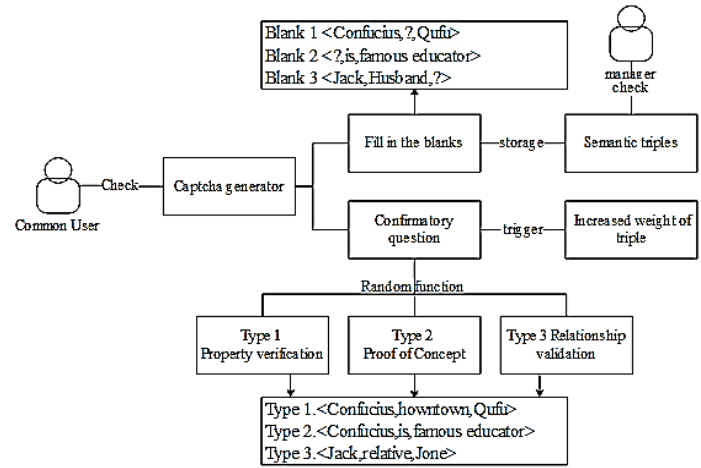

**Fig. 3.** Reverse captcha generation

By calling on the existing network verification code generated interface two type of generated authentication code like figure 4 in the login page, the user has to input verification code instead of traditional numeric verification or image-recognition verification. At last, it will make the weight of generated triples rise in the Confirmatory question or storage new semantic triple in the Fill in the blanks process. All the high score and low score of Top-N triples can be chosen to check again by manager or recirculate in the new round finally. Since the system will review the user's task by generated captchas, if the views are consistent, the credibility will be proved to be high and the score will be increased.

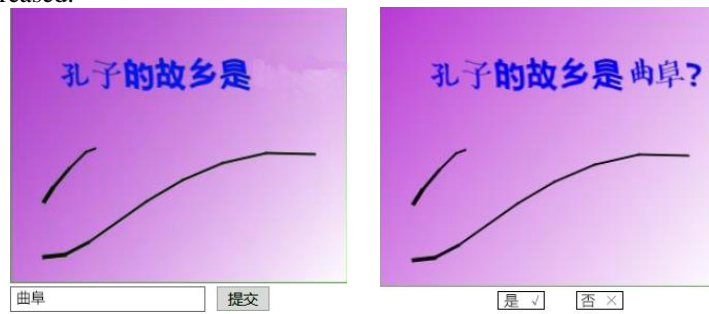

**Fig. 4.** Two type of knowledge Reverse Captcha triple

### 2.5 Different Answer Within Crowdsourcing

During the above two crowdsource task distribution, group task allocation and reverse captcha generation task, may result ambiguity question or irrelevant answer.

For fill-in-blank type of question, we use a relaxation strategy to generate candidate answer. Each answer will allocate a default score at the beginning. By numerable input into the blank, we can extract Top-2 answer generated by the occurrences of each answer, while the lowest answer considered to be irrelevant answer will eliminate after a new cycle.

After the relaxation strategy to eliminate the irrelevant answer, we can add a new type of ambiguity question like "answer A unequal answer B?" used the top-2 answer A and B. If the percentage of "unequal" greater than 35%(self-defined), we regard this task as multi-answer task, both of the top-2 answer will reserve into the knowledge graph. Similarity, If the percentage of "unequal" less than 35%(self-defined), we regard this task as one-answer task, only the top answer will be reserve, and the second-highest answer will auto-reflect to this answer.

## 3 Experimental Evaluation

### 3.1 Design Basic Knowledge graph

To construct a knowledge graph in educational field, we take subject of educational technology (ET) as an example. The basic entity of a subject is course nodes. As the figure 5 depicted, the knowledge graph only exists two kind of nodes, subject node and course node. Relation between then call "course-subject" relation.

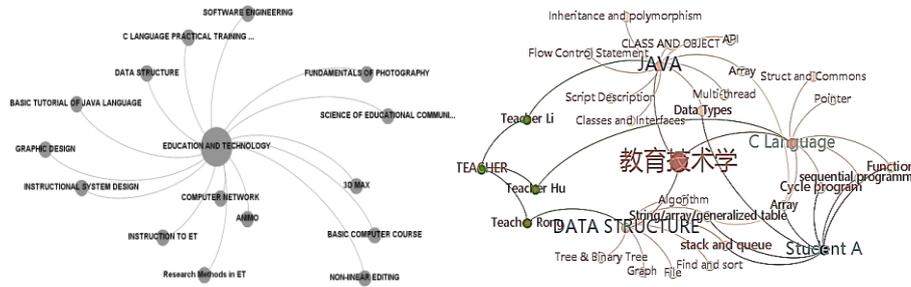

**Fig. 5.** From introduce course nodes into basic KG (left) to all nodes(right)

As Educational entities can divide into school, course, chapter, knowledge and resource, subsequently we introduce the new four kinds of nodes into the basic KG and use different colors to represent the same class to build a distinct knowledge graph, like figure 5(right).

### 3.2 Results Visualization

Based on the node and relationship model built above, we use the visualization plug-in of Echart3.0 to present the construction results in a visual form. As shown in figure 6, the knowledge graph of education domain is composed of schools, courses, knowledge elements, resources and creators.

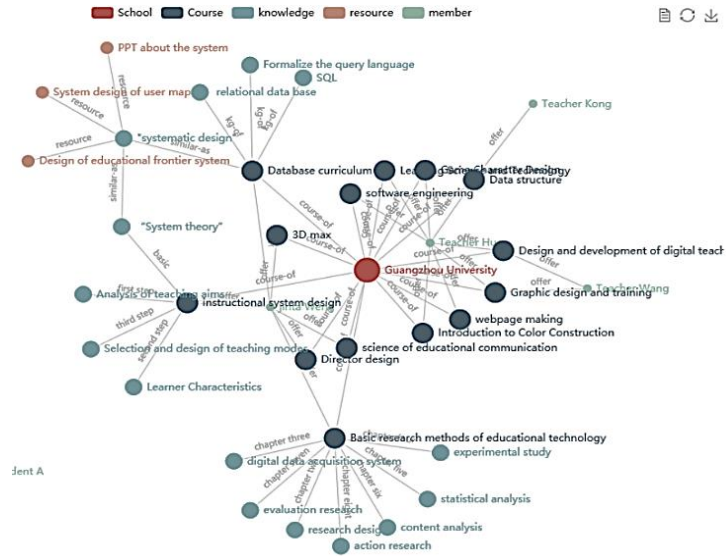

**Fig. 6.** Visualization of the Educational field

### 3.3 Different Subgraph Of KG

By analyzing "teacher" nodes and their offered "course" node, not only we can judge detached teacher and cooperative teacher, but we can also find the teacher ability by the connective nodes. Furthermore, each course nodes can fall into the same category, which make analyses the ability type of teacher possible. We extract the Teacher and their offered course from the KG to formula a subgraph In Figure 7.

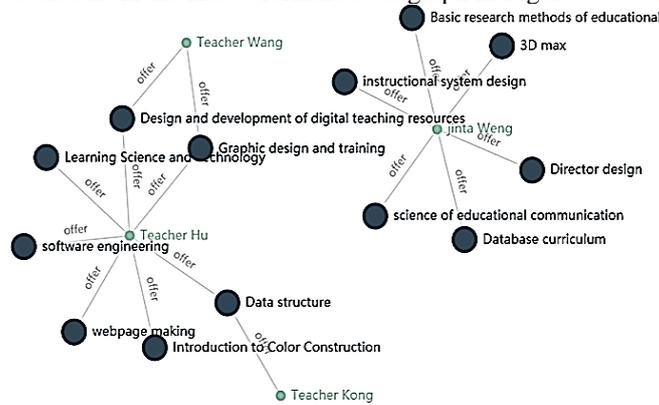

**Fig. 7.** Teacher-Course-Type subgraph

To master and compare the learning situation of different students to locate the poor student and excellent student, we extract the teacher nodes and their offered course from the KG to automatically formula a subgraph as Figure 8 shown.

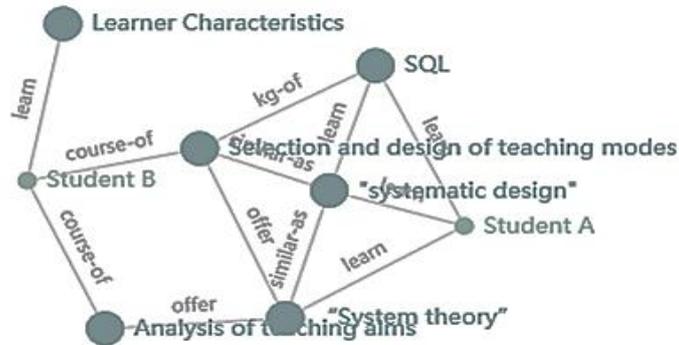

**Fig. 8.** Student-Course-Type subgraph

The third subgraph of Kg are Knowledge-Course-Type subgraph, which make of course node and knowledge-point node. As show in figure 9, the relationship within different course can be associate. By the help of the route from one course node to another node, like the route from "instructional system design" and "database curriculum", we can generate a suitable learning route to student.

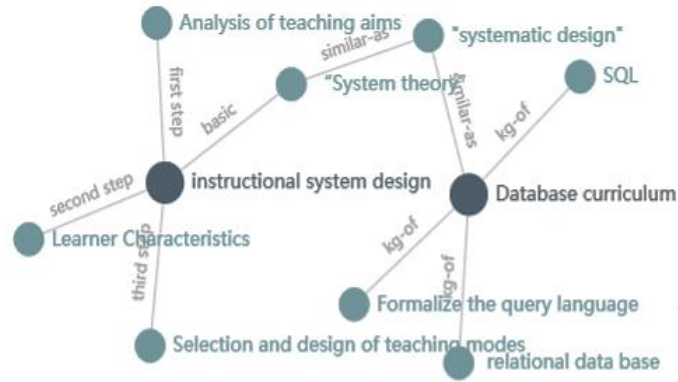

**Fig. 9.** Knowledge-Course-Type subgraph

## 4 Application

### 4.1 Personalized exercises recommendation

Personalized exercises recommendation is an important application of knowledge graph to recommend the most suitable resources for learners. For this application, following factors are generally considered:

1）incremental recommendation: Based on the recent learning topics, it recommends exercises of characteristic knowledge points under specific courses, exercises of related knowledge points under specific courses, exercises of characteristic knowledge points under related courses, and exercises of related knowledge points under related courses. For incremental recommendation, learners' learning situation and curriculum structure developed by teachers will be involved. The nodes involved include students, courses and knowledge. The learning situation of student A is divided into "learned courses" and "unlearned courses", as shown in figure 10.

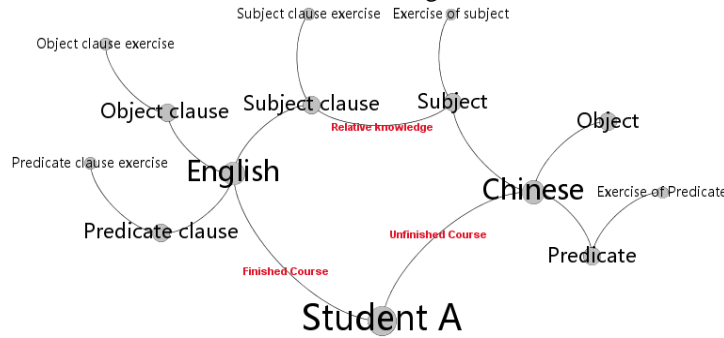

**Fig. 10.** Incremental recommended form of storage

2）Past recommendation: Recommend high error rate of exercises node to learners in the past learning.

For the past recommendation, it will involve entity of students, sub-knowledge and exercises. We put forward an algorithm to recommend the suitable problem set in equation (1) and equation (2). The learner's learning situation LS is reflected by the rate R of finished resources node and the error rate E of exercises node. Rate of resources node graph are the proportion of sum of finished entity within the all resource of same current topic.

In the case that exercises are same error rate, learners who have learned more related resources have lower learning ability than those who have not learned or have learned less about the same number of resources, that is, the former needs to spend more time to complete consolidation exercises. Based on the above conditions, the evaluation algorithm is developed:

$$LS = \frac{1}{R} * E * 100\% \qquad (1)$$

$$R = \frac{\sum_1^{course\ num}(exercise+video+note+other\ resource)}{Number\ of\ all\ resouces} \qquad (2)$$

By setting a certain threshold value P, when the value is greater than the standard threshold value, the default is that the answer is good; when the answer is lower than the threshold value, the answer is considered to be poor. When 20% is set as the comparison threshold in specified topic, the P value above 20% is considered to be more recommended for exercises under this topic, while less than 20% is not recommended. For the determination of contrast threshold, teachers need to make dynamic adjustment according to the average level of students.

### 4.2 Collaborative Creation

Knowledge graph with a good extensible graph structure and human-readable visualization, we develop a knowledge graph tool to allow users add nodes and links to the graph together. Therefore, teachers can carry out group inquiry learning according to the teaching needs, and learners can improve the collected materials according to the topics carried out by teachers.

In this study, we take the ancient poem "Jing Ye Si" as an example and make a preliminary experiment.

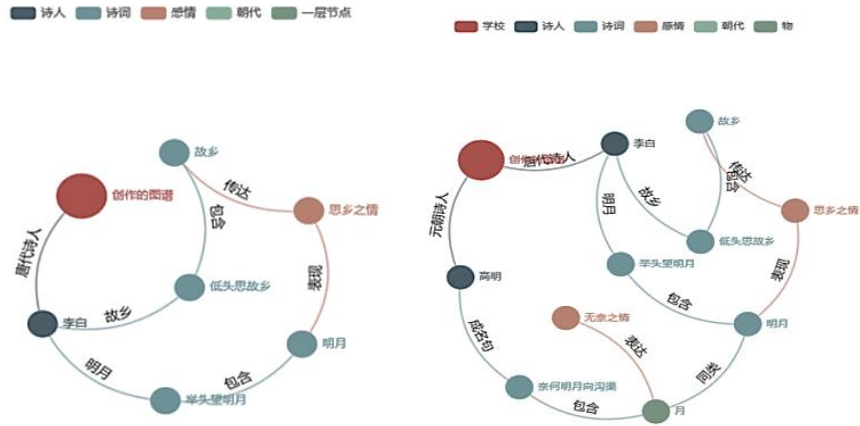

**Fig. 11.** Knowledge graph drawn by user A and user B, left knowledge graph is drawn by User A while right one drawn by User A.

As shown in figure 11 above, the knowledge graph drawn by user A, taking Bai Li's thoughts on A quiet night as an example, can be seen that the author conveys the author's homesickness through the words "bright moon" in "looking up at the bright moon" and "hometown" in "looking down at the hometown".

Basing on A's knowledge graph, student B join and improve the classical knowledge graph. According to the moon in poet "the moon according to ditch" of the author Ming Gao in the yuan dynasty, Ming Gao thought out the moon to convey inner helpless mood. Although the entity moon in this poet within Bai Li poem refers to the same one, it expresses different meaning by the different crowdsourcing user.

To sum up, the cognitive levels of user A and user B are different. Most users can know the "homesick" moon in "Thinking at Quiet Night", but they cannot know the "Helplessness" conveyed by the moon in "Pipa Xing". Therefore, this collaborative creation method can highlight the individual differentiation level under the cognitive level.

## 5 Evaluation and Summary

To certificate the practicality and reliability, we propose two applications and evaluate our knowledge teaching system by 20-point questionnaire. We invite 20 users which make up of 5 teachers and 15 students to ask the question shown in table 1, the average score of questionnaires can fluctuate up and down within 4 points in all dimensions. According to the Richter scale, a score of 4 indicates that respondents have a high sense of system identity. The total average score of the dimensions explored in this study is 3.94, which is approximately equal to 5(highest)*0.8(80%), indicating that the surveyor's satisfaction with the system is close to 80%.

**Table 1 statistical results**

| Problem List | Mean |
|---|---|
| I think this system can help me find the key knowledge points of this subject | 4.00 |
| I think this system can improve my cognitive structure | 3.75 |
| I think this system can conveniently display my learning context (learning portrait) | 3.80 |
| I think this system can facilitate me to master the learning situation of other learners | 3.70 |
| In my opinion, this system can be used as a tool for classroom knowledge network creation and knowledge resource storage | 4.15 |
| I think this system can facilitate teachers' classroom teaching | 3.85 |
| I can use different classification rules to reason on the knowledge graph | 4.20 |
| In my opinion, this knowledge graph can meet the needs of related data search<br>I quickly mastered the use of knowledge graph | 4.15 |
| I can accept the operating interface display of this system | 3.75 |
| I am more interested in the learning method with the help of knowledge graph than the traditional learning method | 3.95 |
| I think the learning method with the help of knowledge graph is better than the traditional learning method | 3.95 |
| 20 Total | 3.94 |

Therefore, we believe that our teaching system based on knowledge graph can solve the shortcomings of traditional adaptive learning system, and teachers and students generally agree with the teaching system based on knowledge graph. However, the above table can also point out the shortcomings of the system: (1) systematic proficiency in question 9 indicates that learners are relatively unfamiliar with the learning mode of knowledge graph;(2) according to question 3, learners do not pay too much attention to the knowledge structure. (3) The teaching practicality of the sixth question highlights the lack of certain teaching practice in this system. Knowledge graph constructed in a crowd-sourced manner requires many users to participate collaboratively, and needs to fully consider the teacher guidance and User mobilization issues.

## Acknowledgment

This work was supported by Guangzhou teaching achievement cultivation project([2017]93)、Guangdong Province Higher Education Teaching Reform Project ([2018]180)